# INTEGRATING AI EDUCATION IN DISCIPLINARY ENGINEERING FIELDS: TOWARDS A SYSTEMS AND CHANGE PERSPECTIVE

**J Schleiss**[1]
Otto von Guericke University Magdeburg
Magdeburg, Germany
https://orcid.org/0009-0006-3967-0492

**A Johri**
George Mason University
Fairfax, USA
https://orcid.org/0000-0001-9018-7574

**S Stober**
Otto von Guericke University Magdeburg
Magdeburg, Germany
http://orcid.org/0000-0002-1717-4133

## ABSTRACT

Building up competencies in working with data and tools of Artificial Intelligence (AI) is becoming more relevant across disciplinary engineering fields. While the adoption of tools for teaching and learning, such as ChatGPT, is garnering significant attention, integration of AI knowledge, competencies, and skills within engineering education is lacking. Building upon existing curriculum change research, this practice paper introduces a systems perspective on integrating AI education within engineering through the lens of a change model. In particular, it identifies core aspects that shape AI adoption on a program level as well as internal and external influences using existing literature and a practical case study. Overall, the paper provides an analysis frame to enhance the understanding of change initiatives and builds the basis for generalizing insights from different initiatives in the adoption of AI in engineering education.

---

[1] *Corresponding Author*
*J Schleiss*
*johannes.schleiss@ovgu.de*

# 1 INTRODUCTION

The use of AI has become increasingly relevant across domains (Broo, Kaynak, and Sait 2022; Patel et al. 2021). The rise of generative AI tools in particular is an example of how the use of AI-based tools, such as ChatGPT, has found high use, especially by students, while their conceptual understanding of how these tools work is limited (Weidener and Fischer 2024; Tsoeu et al. 2023). Thus, the digital transformation accelerated by AI increases the need for AI education in the disciplinary context so that students can make responsible judgments about the use of tools, the outputs they get, and so that they are workforce-ready (Dignum 2021; Cevik Onar et al. 2018).

### *Positioning AI Education*

AI education is a complex undertaking as the field can be positioned from multiple perspectives. First, it describes *teaching about AI* rather than the use of AI tools in teaching (Zawacki-Richter et al. 2019). Second, it can address different target groups, distinguishing between *general, (basic) AI literacy* that builds foundations for the broad public, *domain-specific AI literacy* that aims to build competencies in the context of a disciplinary field, and *expert AI literacy*, which might target Computer Science majors (Schleiss, Laupichler, et al. 2023; Long and Magerko 2020; Laupichler et al. 2022). Third, AI education in engineering education can be seen as a subset of Computer Science (CS) education, similar to the role of programming, which is a basic skill but applicable across disciplines, which requires a greater understanding of how it can be integrated into curricula (Malmi and Johri 2023).

### *Integration of AI Education in Disciplinary Engineering Fields*

With the premise that the use of AI tools and applications will become more dominant across different engineering use cases, it is necessary to address the integration of related AI competencies in the engineering curricula and create systematic research-based understanding. This aligns with the call for interdisciplinarity in engineering education aimed at integrating multiple disciplines to solve a problem (Van den Beemt et al. 2020; Spelt et al. 2009) as well as the continuation of efforts towards digital engineering and use of data and computation in engineering (Cevik Onar et al. 2018).

In the context of engineering education, Kolmos, Hadgraft, and Holgaard (2016) advanced three different curriculum change strategies: first, an *add-on strategy*, referring to including additional courses or components in a curriculum but not changing the overall educational paradigm; second, an *integration strategy*, which involves modifications of programs, and third, a *re-build strategy* which refers to a fundamental change of the educational paradigms of the curricula. These can be used as an analysis lens for the type and depth of curricular change.

The applicability of either of these strategies depends on the context as curricular change does not take place in a vacuum and a range of factors shape curricular reform, especially in interdisciplinary efforts (Knight et al. 2013; Klein 2018). One conceptual framework that helps to understand the factors that influence design decisions on curriculum and course planning is the *Academic Plan Model* (Lattuca and Stark 2009). The model proposes that the development of curricula and courses while being conducted by faculty is also influenced by external and internal forces in a sociocultural context.

In this paper we build upon the three curriculum response strategies proposed by Kolmos, Hadgraft, and Holgaard (2016), the *Academic Plan Model* (Lattuca and Stark 2009), and perspectives from change theory to create practical insights for integrating AI in engineering education. Our approach is guided by the question: *what factors – external, internal, and program level – influence the integrating AI knowledge in engineering education?* We combine findings from literature and a case study of curricular development to provide an analytical frame for integrating AI education in disciplinary engineering fields.

We first review prior work relevant to understanding curriculum change in engineering before moving on to a practical case study that will help us exemplify change by integrating theoretical perspectives. Our move from the concrete to the analytical is deliberate so that the practical insights we generate are theoretically grounded in an integrated manner. Overall, this paper contributes to a system perspective on the integration of AI education in the engineering domain and supports educational strategy development in engineering education.

## 2 RELATED WORK

### 2.1 Transformation towards AI Education in Education

There already exists some work on transformation processes towards AI education that can be used to identify important drivers. Cantú-Ortiz et al. (2020) described a case study of the AI education strategy and programs at the Tecnologico de Monterrey that builds on five strategic initiatives of (1) developing *academic programs*, (2) building up *research capabilities*, (3) *dissemination* through conferences and training seminars, (4) *outreach* through industry-university collaboration and (5) *internationalization* to enhance the exchange. As influencing factors for the program, the paper mentions internationalization and internships, employability, entrepreneurship, academic competitions, growth plans, funding from industry and public sector, AI conferences, and industry partners. Similarly, Southworth et al. (2023) described the "AI Across the Curriculum" initiative at the University of Florida. The initiative is built on pillars of (1) *investment towards compute* and *adding AI-focused faculty*, (2) *developing AI pedagogy* and including *measures* towards it in the quality enhancement plan, and (3) *developing an understanding of AI literacy* in the university and *targeting curriculum development* and development of new academic programs, pathways, research experiences and career development. The complexity and multidisciplinarity of developing AI programs are also highlighted by the experiences of Utrecht University's AI master program (Janssen et al. 2020) and the design of AI education offers for K-12 target groups (Chiu and Chai 2020).

### 2.2 Models for Interdisciplinary Engineering Education

Approaches in interdisciplinary engineering education have the vision to focus on complex real-world problem-solving, the social awareness and entrepreneurial competencies of engineers as well as the improvement of disciplinary programs (Van den Beemt et al. 2020). Thus, related work on change processes and related change efforts do exist, and two recent efforts that can be a good model for AI are the integration of ethics education and sustainability education in engineering curricula. For example, Weiss, Barth, and von Wehrden (2021) analysed 131 case studies of curriculum change towards integrating sustainability and identified six *implementation patterns* ranging from (1) collaborative paradigm change, (2) bottom-up, evolving institutional change, (3) top-down, mandated institutional change, (4) externally driven initiatives, (5) isolated initiatives, and (6) limited institutional change. Also in the context of sustainability, the *curriculum response strategies* of (Kolmos, Hadgraft, and Holgaard 2016) are inspired by responses to sustainable education proposed by (Sterling 2001): making adjustments in the existing system such as improvements or restructuring and changing the educational paradigm as in redesigning the system and institutions. Similarly, Lambrechts (2010) analysed how and to what extent sustainability-related competencies were integrated in Bachelor programs at their university. Next to sustainability, engineering ethics is another topic of curriculum change processes. In this context, Martin, Conlon, and Bowe (2021) analysed literature through four different analytical levels: individual teaching practices, institutional (programs, departments, implementations), policy (accreditation, funding) and culture (paradigms of practice). This suggests that similar to ethics, AI should be integrated across modules, needs to be put into practice and is shaped from different disciplinary domains (Hitt, Holles, and Lefton 2020).

## 3 CASE STUDY

Next to looking at curriculum change from a theoretical perspective, we describe the case of a curricular development of a Bachelor program of AI engineering to understand practical factors of influence in the context of engineering education. The program (210 ECTS (European Transfer Credit System) credits) at the Otto von Guericke University Magdeburg aims at integrating both disciplines, AI and engineering. The interdisciplinary curriculum, situated between AI and engineering, has been developed in a collaborative and participatory process using the method of curriculum workshops (Schleiss, Manukjan, et al. 2023). As a newly developed curriculum, it follows the re-build strategy and recreates educational experiences on a green field. The purpose of the program and vision is to bridge the gap between engineering and AI and train so-called AI engineers who can develop data-driven solutions for engineering use cases. This interdisciplinary skillset focuses on competencies to responsibly work with data and (AI) models as well as having a domain understanding of underlying engineering processes, data and requirements. Moreover, it requires a high level of systematic problem-solving skills and the ability to communicate effectively across disciplines. As a re-build curriculum, about half of the modules are newly developed targeting different educational experiences, for example, projects, flipped-classroom settings, or hackathons. Additionally, a focus lies on developing and using Open Educational Resources (OER) in teaching and collaboration with industry.

The program structure roughly consists of six different elements: fundamental courses, specialisation courses, projects, electives, internship and thesis. The first semesters target the fundamentals in engineering and computer science with a focus on working with data and AI, and related subjects such as maths or working with sensors. After the fundamentals, students select a specialisation of one engineering application area and focus on working with data and AI solutions in this application context. Throughout the program, students have projects that aim to integrate the different knowledge components and bring them into practice. In the higher semesters, the students have three electives allowing for going deeper in some directions. Moreover, in their last semester, students have a 12-week internship and complete the program with a thesis.

The program development was influenced by multiple internal and external drivers. From the external side, AI education is currently high on the political agenda and funding has been provided to establish the study program. Moreover, the content and implementation are shaped through interactions with industry partners. Internally, the program is influenced mostly by the educators who designed the program and secured the funding for its establishment. While acknowledging the need for such a program on the faculty and institutional level, the development was not driven top-down or part of strategic initiatives for further development. Moreover, the program in its form is only possible through the availability of resources, especially concerning faculty staff (mostly financed by external funding) and existing computing resources for the students. The culture at the institution and faculty was initially resisting change, especially concerning novel approaches to teaching and the structure of the curriculum. At the same time, the external funding and governmental support were perceived as validation of the ideas and acceptance of the proposed curriculum. Next, we use this case to exemplify how AI can be integrated into engineering programs and curricula.

## 4 INTEGRATED PERSPECTIVES FOR AI EDUCATION IN ENGINEERING FIELDS

### 4.1 Academic Plan Model for AI Education in Engineering

To move towards a system perspective on integrating AI education in disciplinary engineering fields, we integrate the findings from theory and practice towards a system model that highlights external and internal influences. The abstract model (Figure 1) builds upon the *Academic Plan Model* (Lattuca and Stark 2009) and frames the context that is used to understand and describe how change might occur towards adopting AI education. In the following, we describe the different contexts in more detail and highlight how these behave specific to AI education.

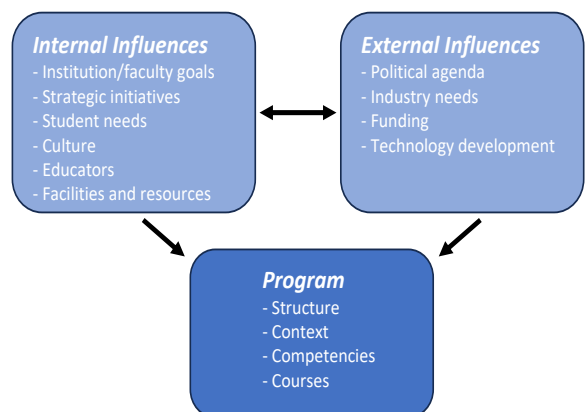

*Figure 1. System perspective on influences of curriculum change inspired by (Lattuca and Stark 2009)*

### 4.1.1. Program Level for AI Education in Engineering

Within the context of an educational environment, the central element is the *program* with its structure, context, competencies, and content such as instructional resources, processes and assessment and evaluation. A program includes content aspects (curriculum) and organizational aspects (regulations). A curriculum is usually described through the goal, purpose, content and sequence of how different modules work together. Different modules might consist of one or more courses and include instructional resources, processes, assessment and evaluation, all targeted to build out competencies for learners. As introduced earlier, when looking at changes in curricula, response strategies can be roughly distinguished into an add-on, integration and re-build strategy (Kolmos, Hadgraft, and Holgaard 2016). Thus, we can use these perspectives for integrating AI in engineering education.

Following this structure, it becomes clear that we need to understand how to develop (interdisciplinary) curricula that integrate AI and different engineering domains including aspects of what competencies and content are relevant and in what sequence it should be taught. This can lead to adding new competencies, integrating them into existing curricula and courses or re-building them. More specifically, from a course perspective, there is the need to understand what our learners need in their interaction with AI, what relevant instructional resources are, how we can structure instructional processes and how we can assess and evaluate the learning.

A few curricula and course development efforts exist in the literature. Work in the context of curricula, development has been focused on the idea of multi- and interdisciplinary design highlighting the diverse nature of AI (Janssen et al. 2020; Southworth et al. 2023). The research on understanding relevant AI competencies is still in development (Tenório and Romeike 2024; Almatrafi, Johri, and Lee 2024; Wolters, Arz Von Straussenburg, and Riehle 2024). Regarding course development, there have been efforts to provide structure for educators, e.g. the AI course design planning framework (Schleiss, Laupichler, et al. 2023). Moreover, there have been multiple efforts in the definition of AI literacy assessment instruments (Hornberger, Bewersdorff, and Nerdel 2023; Laupichler, Aster, and Raupach 2023).

***Exemplary Core Questions***

- What is the purpose of integrating AI in the program?
- What AI competencies are targeted and should be integrated?
- Does the adoption of curricula or courses towards AI follow an add-on, integration and re-build strategy?
- What learning activities and assessment strategies are suitable for AI education?

### 4.1.2. Internal Influences on AI Education in Engineering

Within the socio-cultural context, we find internal and external forces. Concerning internal influences Lattuca and Stark (2009) distinguish between institutional such as the objectives of the institution, resources or governance, and unit-level influences such as educators in faculty, discipline culture and student characteristics. From the perspective of adopting AI education in engineering education, it is important to understand internal drivers in the institution. These can be roughly distinguished as focusing on people and focusing on organizational structures. People are central to change, especially in education. In adopting AI education, one key challenge is the readiness of faculty for change, be it the level of expertise and skills in AI, the needed support structures, or the potential resistance to change (Stark et al. 1988). Moreover, aspects such as governance structures are important to consider. For example, Knight et al. (2013) highlighted the importance of organizational features, such as the appointment of a program director and faculty appointment within the interdisciplinary field. Additionally, in the context of AI, computational resources are needed for deployment, experiments and teaching. Thus it is important to consider what resources are available on a faculty and institutional level.

*Exemplary Core Questions*

- Who is driving the integration of AI education? Is it led by leadership (top-down) or by faculty/educators/students (bottom-up)?
- How well embedded is AI education in the faculty goals or strategic initiatives?
- How well-equipped are educators to teach AI topics?
- What governance structures can support the integration of AI in curricula?
- How is the funding and resource situation, e.g. for compute at the institution?

### 4.1.3. External Influences on AI Education in Engineering

Next to internal influences, the educational environment is influenced by external forces. These can include market forces, political agenda and initiatives, accrediting agencies, disciplinary associations, technology advancements, funding or industry needs. Concerning AI education, we can highlight five major forces from theory and practice. First, AI development has mostly accelerated due to technological advancements in computing and model architectures in recent years, leading to new market forces and industry demand for AI talent. This creates a demand for more AI education at universities. Second, the need for AI talent and research is also picked up by governments and supported through funding or political initiatives, which create an opportunity for universities to create new programs or research initiatives. Third, new competencies are also picked up by accrediting bodies, even though a recent article highlighted that currently AI is not yet broadly integrated into engineering accreditation (Tsoeu et al. 2023). Fourth, the use of AI systems has ethical, legal and social implications and is also a topic of regulatory aspects. Regulations like the EU AI Act influence the possible use of AI systems but in a broader sense also what should be taught in the context of AI education to educate responsible engineers. Fifth, there exists the challenge of a high volatility of skills due to technology advancements. Identifying the relevant skills and adopting courses and curricula at the required speed remains challenging for higher education.

*Exemplary Core Questions*

- How important is the topic of AI considered on the current political agenda?
- What is the need for AI skills in the industry? Who are potential partners?
- How much funding is available to develop new programs?
- How is the constant technology change in AI influencing relevant skills, adoption in programs and in courses?

### 4.2 Change Theory Perspective for AI Education in Engineering

Curriculum change can be viewed from three dimensions: (1) *triggers* of change, (2) *drivers and barriers* of change, and (3) the *type and content* of change. *Triggers of change* refer to the impetus of change, from forces within the institution or from external sources (Lattuca and Stark 2009), and it can be normative or goal-oriented (Fumasoli and Lepori 2011). *Drivers and barriers of change* can again be external or internal influences (Gruba et al. 2004). Internally, questions about the institutional goals, resource availability, governance structures, and readiness of faculty for integrating and changing (Gruba et al. 2004; Roberts 2015). Moreover, student needs and interests as well as their characteristics and abilities might drive or hinder change (Gruba et al. 2004). Externally, the market forces, the political agenda, accreditation agencies or disciplinary associations or change in other institutions might influence the change process (Gruba et al. 2004). Finally, the *type and content of change* vary and can be linked to different response strategies, e.g. adding new degrees or course work, new subjects or integrating aspects in core or electives (Gruba et al. 2004; Kolmos, Hadgraft, and Holgaard 2016). From a content perspective, there is the question of what competencies are relevant and categorizing their integration, e.g. in vertical integration (explicit mentioning of competency), horizontal integration (competency implicitly integrated across the curriculum), and a combination of both (Lambrechts 2010). For an overview, the system and the change perspective are summarized in Table 1.

*Table 1. Integrated Systems and Change Perspective towards integrating AI education in EE*

| Change Dimensions | System Elements | | Program Level Changes for integrating AI education in engineering education |
|---|---|---|---|
| | *Internal* | *External* | |
| *Trigger* | Faculty interests and motivation<br>Strategic initiatives from leadership<br>Student demands for integration of AI | Accreditation processes<br>Funding from industry and government<br>Market demands for AI graduates<br>Risks of the use of AI | Awareness and training in AI for faculty and leadership as well as new faculty hires can trigger change<br>Loop between market demands and student needs for AI related competencies and new learning experiences<br>Responsible and ethical use of AI is key for engineers |
| *Drivers and Barriers* | Institutional goals<br>Available resources<br>Governance structure<br>Readiness of faculty | Accreditation processes<br>Funding from industry and government<br>Internationalization<br>Exchange with others<br>Change in other HEIs | Faculty AI skills, motivation and time as well as related governance structures relevant for program level change<br>Funding and accreditation accelerates the change process towards integrating AI in EE<br>AI technology advancement driven through industry - Industry collaboration more important<br>Student demands accelerates AI adoption |
| *Type and Content* | Curricular regulations<br>Agility of faculty administration | Availability of competence frameworks<br>Accreditation | Identification of relevant competencies and their integration across curricula as main challenge<br>Selecting adequate change response strategy and integration strategy and related teaching approaches |

## 5. DISCUSSION AND IMPLICATIONS

We have introduced a system perspective on integrating AI education in disciplinary engineering fields building upon literature and a practical case study. Connecting to other transformation processes, this work contributes towards an improved understanding of the required change processes for integrating AI competencies in disciplinary engineering fields. It can serve as a thinking frame for educators and faculty leaders to see where they stand and how to develop a change strategy towards integrating AI into their educational offers. We recognize that the paper has certain limitations. First, the system perspective as an analysis frame is inductively developed from a case study and literature only and not yet validated in use. Moreover, the selection of the case might have biased the development. These limitations can be overcome in future through analysis of further use cases and integration of expert feedback, in particular with comparison to previous integration efforts of other cross-ranging aspects such as sustainability and ethics. From the analysis frame, we can also identify implications for further research. First, more case studies should be collected and analysed within the analysis frame to build a foundation for transferring local best practices to more general contexts. Second, more research is needed to understand program structures and ways to integrate AI competencies in engineering education in a disciplinary and interdisciplinary way. Third, internal and external drivers of AI education can benefit from more research, in particular to map change theories to the drivers (Reinholz and Andrews 2020). Fourth, the analysis frame could be further enhanced with metrics, e.g. a readiness scale, to support the change processes at universities and allow for comparisons on a faculty, institutional, national or international level.

**Acknowledgements**
This work is partly supported by the German Federal Ministry of Education and Research under grant number 16DHBKI008 and U.S. NSF Awards# 2319137, 204863, USDA/NIFA Award#2021-67021-35329. Any opinions, findings, and conclusions or recommendations expressed in this material are those of the authors and do not necessarily reflect the views of the funding agencies.

**REFERENCES**

Almatrafi, Omaima, Aditya Johri, and Hyuna Lee. 2024. "A Systematic Review of AI Literacy Conceptualization, Constructs, and Implementation and Assessment Efforts (2019-2023)." *Computers and Education Open*, 100173.

Broo, Didem Gürdür, Okyay Kaynak, and Sadiq M Sait. 2022. "Rethinking Engineering Education at the Age of Industry 5.0." *Journal of Industrial Information Integration* 25:100311.

Cevik Onar, Sezi, Alp Ustundag, Çigdem Kadaifci, and Basar Oztaysi. 2018. "The Changing Role of Engineering Education in Industry 4.0 Era." In *Industry 4.0: Managing The Digital Transformation*, edited by Alp Ustundag and Emre Cevikcan, 137–51. Cham: Springer International Publishing. https://doi.org/10.1007/978-3-319-57870-5_8.

Chiu, Thomas K. F., and Ching-sing Chai. 2020. "Sustainable Curriculum Planning for Artificial Intelligence Education: A Self-Determination Theory Perspective." *Sustainability* 12 (14): 5568. https://doi.org/10.3390/su12145568.

Dignum, Virginia. 2021. "The role and challenges of education for responsible AI." *London Review of Education* 19 (1). https://doi.org/10.14324/LRE.19.1.01.

Fumasoli, Tatiana, and Benedetto Lepori. 2011. "Patterns of Strategies in Swiss Higher Education Institutions." *Higher Education* 61 (2): 157–78. https://doi.org/10.1007/s10734-010-9330-x.

Gruba, Paul, Alistair Moffat, Harald Søndergaard, and Justin Zobel. 2004. "What Drives Curriculum Change?" In *Proceedings of the Sixth Australasian Conference on Computing Education*. Vol. 30.

Hitt, Sarah Jayne, Cortney E. P. Holles, and Toni Lefton. 2020. "Integrating Ethics in Engineering Education through Multidisciplinary Synthesis, Collaboration, and Reflective Portfolios." *Advances in Engineering Education*. https://eric.ed.gov/?id=EJ1279776.

Hornberger, Marie, Arne Bewersdorff, and Claudia Nerdel. 2023. "What Do University Students Know about Artificial Intelligence? Development and Validation of an AI Literacy Test." *Computers and Education: Artificial Intelligence* 5 (January):100165. https://doi.org/10.1016/j.caeai.2023.100165.

Janssen, Christian P., Rick Nouwen, Krista Overvliet, Frans Adriaans, Sjoerd Stuit, Tejaswini Deoskar, and Ben Harvey. 2020. "Multidisciplinary and Interdisciplinary Teaching in the Utrecht AI Program: Why and How?" *IEEE Pervasive Computing* 19 (2): 63–68. https://doi.org/10.1109/MPRV.2020.2977741.

Klein, Julie Thompson. 2018. "Current Drivers of Interdisciplinarity: The What and the Why." In *Promoting Interdisciplinarity in Knowledge Generation and*

*Problem Solving*, 14–28. IGI Global. https://doi.org/10.4018/978-1-5225-3878-3.ch002.

Knight, David B., Lisa R. Lattuca, Ezekiel W. Kimball, and Robert D. Reason. 2013. "Understanding Interdisciplinarity: Curricular and Organizational Features of Undergraduate Interdisciplinary Programs." *Innovative Higher Education* 38 (2): 143–58. https://doi.org/10.1007/s10755-012-9232-1.

Kolmos, Anette, Roger G. Hadgraft, and Jette Egelund Holgaard. 2016. "Response Strategies for Curriculum Change in Engineering." *International Journal of Technology and Design Education* 26 (3): 391–411. https://doi.org/10.1007/s10798-015-9319-y.

Lambrechts, Wim. 2010. "The Integration of Sustainability in Competence Based Higher Education." In *ERSCP-EMSU Conference*.

Lattuca, Lisa R., and Joan S. Stark. 2009. *Shaping the College Curriculum: Academic Plans in Context*. John Wiley & Sons.

Laupichler, Matthias Carl, Alexandra Aster, and Tobias Raupach. 2023. "Delphi Study for the Development and Preliminary Validation of an Item Set for the Assessment of Non-Experts' AI Literacy." *Computers and Education: Artificial Intelligence* 4 (January):100126. https://doi.org/10.1016/j.caeai.2023.100126.

Laupichler, Matthias Carl, Alexandra Aster, Jana Schirch, and Tobias Raupach. 2022. "Artificial Intelligence Literacy in Higher and Adult Education: A Scoping Literature Review." *Computers and Education: Artificial Intelligence* 3:100101. https://doi.org/10.1016/j.caeai.2022.100101.

Long, Duri, and Brian Magerko. 2020. "What Is AI Literacy? Competencies and Design Considerations." In *Proceedings of the 2020 CHI Conference on Human Factors in Computing Systems*, 1–16. Honolulu HI USA: ACM. https://doi.org/10.1145/3313831.3376727.

Malmi, Lauri, and Aditya Johri. 2023. "A Selective Review of Computing Education Research." *International Handbook of Engineering Education Research*, 573–93.

Martin, Diana Adela, Eddie Conlon, and Brian Bowe. 2021. "A Multi-Level Review of Engineering Ethics Education: Towards a Socio-Technical Orientation of Engineering Education for Ethics." *Science and Engineering Ethics* 27 (5): 60. https://doi.org/10.1007/s11948-021-00333-6.

Patel, Amit R, Kashyap K Ramaiya, Chandrakant V Bhatia, Hetalkumar N Shah, and Sanket N Bhavsar. 2021. "Artificial Intelligence: Prospect in Mechanical Engineering Field—a Review." *Data Science and Intelligent Applications: Proceedings of ICDSIA 2020*, 267–82.

Reinholz, Daniel L., and Tessa C. Andrews. 2020. "Change Theory and Theory of Change: What's the Difference Anyway?" *International Journal of STEM Education* 7 (1): 1–12. https://doi.org/10.1186/s40594-020-0202-3.

Roberts, Pamela. 2015. "Higher Education Curriculum Orientations and the Implications for Institutional Curriculum Change." *Teaching in Higher Education* 20 (5): 542–55. https://doi.org/10.1080/13562517.2015.1036731.

Schleiss, Johannes, Matthias Carl Laupichler, Tobias Raupach, and Sebastian Stober. 2023. "AI Course Design Planning Framework: Developing Domain-Specific AI Education Courses." *Education Sciences* 13 (9): 954. https://doi.org/10.3390/educsci13090954.

Schleiss, Johannes, Anke Manukjan, Michelle Ines Bieber, Philipp Pohlenz, and Sebastian Stober. 2023. "Curriculum Workshops As A Method Of Interdisciplinary Curriculum Development: A Case Study For Artificial Intelligence In Engineering." https://doi.org/10.21427/XTAE-AS48.

Southworth, Jane, Kati Migliaccio, Joe Glover, Ja'Net Glover, David Reed, Christopher McCarty, Joel Brendemuhl, and Aaron Thomas. 2023. "Developing a Model for AI Across the Curriculum: Transforming the Higher Education Landscape via Innovation in AI Literacy." *Computers and Education: Artificial Intelligence* 4:100127. https://doi.org/10.1016/j.caeai.2023.100127.

Spelt, Elisabeth J. H., Harm J. A. Biemans, Hilde Tobi, Pieternel A. Luning, and Martin Mulder. 2009. "Teaching and Learning in Interdisciplinary Higher Education: A Systematic Review." *Educational Psychology Review* 21 (4): 365–78. https://doi.org/10.1007/s10648-009-9113-z.

Stark, Joan S., Malcolm A. Lowther, Michael P. Ryan, and Michele Genthon. 1988. "Faculty Reflect on Course Planning." *Research in Higher Education* 29 (3): 219–40. https://doi.org/10.1007/BF00992924.

Sterling, Stephen. 2001. "Sustainable Education: Re-Visioning Learning and Change. Schumacher Briefings."

Tenório, Kamilla, and Ralf Romeike. 2024. "AI Competencies for Non-Computer Science Students in Undergraduate Education: Towards a Competency Framework." In *Proceedings of the 23rd Koli Calling International Conference on Computing Education Research*, 1–12. Koli Calling '23. New York, NY, USA: Association for Computing Machinery. https://doi.org/10.1145/3631802.3631829.

Tsoeu, Mohohlo, Rendani Maladzi, Nomcebo Mthombeni, Katleho Moloi, Tebogo Mashifana, and Fulufhelo Nemavhola. 2023. "Engineering, the Profession in Trouble: Lack of Programme Development Standards That Support the AI Chatbot? A System View." In *2023 World Engineering Education Forum - Global Engineering Deans Council (WEEF-GEDC)*, 1–6. https://doi.org/10.1109/WEEF-GEDC59520.2023.10343985.

Van den Beemt, Antoine, Miles MacLeod, Jan Van der Veen, Anne Van de Ven, Sophie van Baalen, Renate Klaassen, and Mieke Boon. 2020. "Interdisciplinary Engineering Education: A Review of Vision, Teaching, and Support." *Journal of Engineering Education* 109 (3): 508–55. https://doi.org/10.1002/jee.20347.

Weidener, Lukas, and Michael Fischer. 2024. "Artificial Intelligence in Medicine: Cross-Sectional Study Among Medical Students on Application, Education, and Ethical Aspects." *JMIR Medical Education* 10 (1): e51247. https://doi.org/10.2196/51247.

Weiss, Marie, Matthias Barth, and Henrik von Wehrden. 2021. "The Patterns of Curriculum Change Processes That Embed Sustainability in Higher Education

Institutions." *Sustainability Science* 16 (5): 1579–93. https://doi.org/10.1007/s11625-021-00984-1.

Wolters, Anna, Arnold F Arz Von Straussenburg, and Dennis M Riehle. 2024. "AI Literacy in Adult Education-A Literature Review."

Zawacki-Richter, Olaf, Victoria I. Marín, Melissa Bond, and Franziska Gouverneur. 2019. "Systematic Review of Research on Artificial Intelligence Applications in Higher Education – Where Are the Educators?" *International Journal of Educational Technology in Higher Education* 16 (1): 39. https://doi.org/10.1186/s41239-019-0171-0.